\documentstyle[twocolumn,floats,prb,aps,psfig]{revtex}

\begin{document}
\newcommand{\degree}{$^{\rm\circ} $}
\newcommand{\pcite}{\protect\cite}
\newcommand{\pref}{\protect\ref}

\title{Comparative Bending Dynamics in DNA with and without A-tracts}

\author{Alexey K. Mazur and Dimitri E. Kamashev}

\address{Laboratoire de Biochimie Th\'eorique, CNRS UPR9080\\
Institut de Biologie Physico-Chimique\\
13, rue Pierre et Marie Curie, Paris,75005, France.\\
FAX:+33[0]1.58.41.50.26. Email: alexey@ibpc.fr}

\maketitle

\begin{abstract}

The macroscopic curvature of double helical DNA induced by regularly
repeated adenine tracts is well-known but still puzzling. Its physical
origin remains controversial even though it is perhaps the
best-documented sequence modulation of DNA structure.  We report here
the results of comparative theoretical and experimental studies of
bending dynamics in 35-mer DNA fragments. This length appears large
enough for the curvature to be distinguished by gel electrophoresis.
Two DNA fragments, with identical base pair composition, but different
sequences are compared. In the first one, a single A-tract motif was
four times repeated in phase with the helical screw whereas the second
sequence was "random". Both calculations and experiments indicate that
the A-tract DNA is distinguished by the large static curvature and
characteristic bending dynamics, suggesting that the computed effect
corresponds to the experimental phenomenon. The results poorly agree
with earlier views that attributed a decisive role in DNA bending to
sequence specific base pair stacking or binding of solvent
counterions, but lend additional support to the hypothesis of a
compressed frustrated state of the backbone as the principal physical
cause of the static curvature. We discuss the possible ways of
experimental verification of this hypothesis.  \end{abstract}

\section*{Introduction}

It is generally accepted that the base pair sequence can affect the
overall form of the DNA double helix. Intrinsic DNA bending is the
simplest such effect.  Natural static curvature was discovered nearly
twenty years ago in DNA containing regular repeats of $\rm A_nT_m,\
with\ n+m>3$, called A-tracts \cite{Marini:82,Wu:84,Hagerman:84}.
Since then this intriguing phenomenon has been actively studied, with
several profound reviews of the results published in different years
\cite{Diekmann:87a,Hagerman:90,Crothers:90,Crothers:92,Olson:96,Crothers:99}.
It is known that the curvature is directed towards the minor grooves
of A-tracts and/or the major grooves of the junction zones between
them, and that its magnitude is around 18\degree\ per A-tract.
However, the exact sites and the character of local bends remain a
matter of debate as well as their mechanism and physical origin.

Already the pioneering conformational calculations of the seventies
showed that the DNA double helix exhibits significant bendability
which is anisotropic and sequence dependent
\cite{Namoradze:77,Zhurkin:79}. Based upon these views the wedge model
offered the very first explanation of bending induced by A-tracts by
postulating that stacking in ApA steps is intrinsically non-parallel
\cite{Trifonov:80}. Modified versions of this theory accounted for a
substantial part of available experimental data, with good scores of
curvature prediction from sets of fitted wedge angles
\cite{DeSantis:90,Bolshoy:91,Liu:01}. At the same time, clear
experimental counter-examples exist where bending could not result
from simple accumulation of wedges \cite{Dlakic:96a,Dlakic:98}. The
junction model \cite{Levene:83,Wu:84} better than other theories explained
experimental data on gel retardation of curved DNA. It originated from
an idea that a bend should occur when two different DNA forms are
stacked \cite{Selsing:79}. If poly-{dA} double helix had a special B'
form as suggested by some data \cite{Alexeev:87} the helical axis
should be kinked when an A-tract is interrupted by a random sequence.
In turn, the X-ray data are best interpreted with an alternative
theory that postulates that bending is intrinsic in most DNA sequences
except A-tracts which are straight
\cite{Calladine:88,Maroun:88,Dickerson:94}. Another interesting model
attracted attention in the recent years, namely, bending by
electrostatic forces that result from neutralization of phosphates by
solvent cations trapped in the minor grooves of A-tracts
\cite{Hud:99}. This problem is of general importance because the
accumulated large volume of apparently paradoxical observations
suggests that some essential features are still unknown that may be
essential for the fine structure and the biological function of the
DNA molecule.

One of us has recently proposed a new hypothesis of the physical
origin of intrinsic bends in double helical DNA \cite{Mzjacs:00}.
According to it, the sugar-phosphate backbone in physiological
conditions is slightly compressed, that is the equilibrium specific
length of the corresponding free polymer in the same solvent is larger
than that in the canonical B-form. Therefore, the backbone ``pushes''
stacked base pairs, forcing them to increase the helical twist and
rise while the stacking interactions oppose this. As a result, the
backbone increases its length by deviating from its regular spiral
trace and wanders along the helical surface causing quasi-sinusoidal
modulations of DNA grooves. Concomitant base stacking perturbations
result in macroscopic static curvature when certain properties of base
pairs alternate along the sequence in phase with the helical screw.

Drew and Travers (1984, 1985) apparently were the first to notice that
narrowing of both DNA grooves at the inner edge of a bend is a
necessary and sufficient condition of bending, and that an unusual
local groove width should be accompanied by structural perturbations
beyond this region. They, and later Burkhoff and Tullius (1987),
considered the preference of narrow and wide minor groove profiles by
certain sequences as the possible original cause of this effect.
Sprous et al (1999) proposed a similar idea within the context of the
junction model. In a certain sense, the compressed backbone theory
continued the same line of thinking. Unlike other models, it naturally
explains well-known environmental effect upon the A-tract curvature,
notably, its reduction with temperature
\cite{Diekmann:85,Diekmann:87c,Jerkovic:00} and addition of
dehydrating agents \cite{Marini:84,Sprous:95,Dlakic:96b} (see
discussion in Mazur (2000)). This theory certainly needs further
examination and critical comparison with other models in both
calculations and experiments.

Conformational modeling earlier helped to shed light upon many aspects
of the above problems. Construction of spatial DNA traces from local
wedge parameters combined with Monte Carlo simulations of loop closure
was applied to check different hypotheses and to estimate local
bending angles from experimental data \cite{Levene:83,Koo:90}. Energy
calculations revealed that bending may be easier at some dinucleotide
steps and in certain specific directions \cite{Zhurkin:79,Ulyanov:84},
with experimental sequence effects reproduced in some remarkable
examples \cite{Sanghani:96}. DNA was shown to have local energy minima
in bent conformations corresponding to the junction model
\cite{Kitzing:87,Chuprina:88}. All atom Monte Carlo calculations
showed that narrowing of the A-tract minor groove with a few
NMR-derived restraints may be sufficient to provoke the curvature
\cite{Zhurkin:91}.

The simplest set-up for modeling DNA bending is to take a straight
symmetrical double helix and let it bend spontaneously with no extra
forces applied, that is due to generic atom-atom interactions.  This
``naive'' approach has recently become possible owing to the progress
in methodology of molecular dynamics (MD) calculations of nucleic
acids \cite{Cheatham:00}, which was demonstrated by successful
simulations of several curved and straight DNA fragments in realistic
environment including explicit water and counterions
\cite{Young:98,Sprous:99}. The character of bending qualitatively
agreed with the theories outlined above so that none of them could be
preferred. Thorough discriminative testing would require more
extensive sampling of bending events, which should become possible in
future. Detailed structures of short A-tract fragments have also been
studied by MD \cite{Sherer:99,Strahs:00}.

The major obstacle in free MD simulations of intrinsic curvature is
the limited capacity of sampling of bending events. The physical time
of transition between straight and bent conformations may be too long
for a statistically significant number of such events to be
accumulated in simulations. Moreover, experimental effects may not
appear during infinitely long MD because models are never perfect.  To
circumvent these difficulties, we employed a different strategy. We
first looked for, and found a short A-tract motif that could
reproducibly induce stable bends in DNA during a few nanoseconds of MD
with a simplified model of B-DNA.  We used this motif to construct
longer double helices with intrinsic curvature {\em in silico} and we
could increase the length of DNA fragments in calculations to 35 bp,
which makes possible a direct comparison with experiments {\em in
vitro}.

The two 35-mer DNA fragments we study here have identical base pair
composition and differ only by their sequences. The first fragment is
the designed A-tract repeat while the other sequence is ``random''.
All MD trajectories start from canonical straight A- and B-DNA
conformations.  For the A-tract DNA fragment they converged to a
single statically bent state with planar curvature towards the
narrowed minor grooves at 3' ends of A-tracts. The magnitude of
bending is close to the experimental estimates.  The random fragment
was not straight as well, but its curvature was much less significant
and less planar. In gel migration assays the two molecules produce
well-resolved distinct bands, with the A-tract sequence demonstrating
a reduced mobility characteristic of curved DNA. These results suggest
that the intrinsic DNA curvature reproduced in calculations
corresponds to the experimental phenomenon. The bending dynamics
qualitatively agrees with the compressed backbone theory, but it cannot
be accounted for by other models.

\section*{Materials and Methods}

\subsection*{Calculations}

Molecular dynamics simulations have been performed by the internal
coordinate method (ICMD) \cite{Mzjcc:97,Mzbook:01} including special
technique for flexible sugar rings \cite{Mzjchp:99}, with AMBER94
\cite{AMBER94:,Cheatham:99} force field and TIP3P water \cite{TIP3P:}.
All calculations were carried out without cut-offs and boundary
conditions. The time step was 10 fsec. The so-called minimal model of
B-DNA was used \cite{Mzjacs:98,Mzjcc:01}. It includes only a partial
hydration shell and treats counterion and long range solvation effects
implicitly. Advantages as well as limitations of this approach have
been reviewed elsewhere \cite{Cheatham:00}. The model has no other
bias towards bent or non-bent conformations except the base pair
sequence.

The starting fiber A- and B-DNA models were constructed
from the published atom coordinates \cite{Arnott:72}. The hydration
protocols were same as before, \cite{Mzjacs:00} with an identical
number of explicit water molecules in A- and B-DNA starts. Programs
Curves, \cite{Curves:} XmMol, \cite{XmMol:} and Mathematica by Wolfram
Research Inc. were employed for graphics and data analysis.

\begin{figure}
\centerline{\psfig{figure=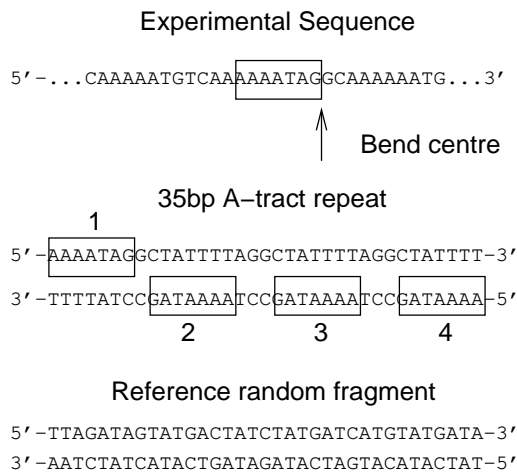,width=8cm,angle=0.}}
\caption{\label{Falseq}
Construction of 35 bp double stranded DNA fragments. The top sequence
with the boxed heptamer motif $\rm AAAATAG$ is taken from the
trypanosome kinetoplast DNA \pcite{Wu:84}. The A-tracts are numbered
and their centers are separated by approximately 10 bp. The reference
random fragment has the same base pair content as the 35-mer repeat,
but its sequence has been manually re-shuffled to exclude any A-tract
motives.}\end{figure}

The two 35pb DNA fragments are referred to below as At and nAt, for
the A-tract repeat and the non-A-tract DNA, respectively. For both
fragments two long MD trajectories were computed starting from either
A- or B- canonical DNA forms. These four trajectories are referred to
as At-A, At-B, nAt-A, and nAt-B, respectively, where the last
character indicates the starting state. All trajectories were
continued to 20 ns except At-B which was stopped at about 12 ns
because it had clearly converged long before.

\subsection*{Oligonucleotides and construction of 5'-labeled DNA
probes}

The sequences of the 35 nt-long synthetic oligonucleotides used here
are shown in Fig. \ref{Falseq}. The double stranded DNAs were obtained
by annealing of the two complementary oligonucleotides, one of them
labeled with T4 polynucleotide kinase and [$^{32}$-P]-ATP. The
annealing was carried out by incubating the oligonucleotides (300 nM)
for 3 min at 80\degree C in 20 mM Tris-HCl (pH 8.0), 400 mM NaCl, 0.2
mM EDTA and then allowing them to cool slowly.

\subsection*{Gel mobility assays}

Mobility of the DNA fragments was analyzed in 16\% gels (acrylamide to
bis-acrylamide, 29:1) buffered with 90 mM Tris-borate, 1 mM EDTA, pH
8.6. Gels were pre-run under constant power until stabilization of
current. End-labeled DNA in a buffer containing 20 mM Tris-HCL, 50 mM
NaCL, 7\% glycerol, pH 8.0 and bromophenol-blue was loaded onto the
gel. The electrophoresis was performed under constant power and
constant temperature of 8\degree C. The dried gels were exposed to
storage phosphor screens and visualized on a 400S PhosphorImager
(Molecular Dynamics).

\section*{Results and Discussion}

\subsection*{Construction of DNA Fragments}

Figure \ref{Falseq} explains how the two DNA fragments used in our
study have been constructed. The A-tract motif $\rm AAAATAG$
originally attracted our attention in MD simulations of the natural
DNA shown in Fig. \ref{Falseq} \cite{Mzjbsd:01}, which is the first
curved DNA locus studied {\em in vitro} \cite{Wu:84}.  The 35 bp
A-tract fragment was constructed by repeating this motif four times
and it had to be inverted to make the two DNA termini symmetrical.
Such inversion should not affect bending, \cite{Koo:86} but is
essential for simulations because the 3'- and 5'-end A-tracts may
represent qualitatively different boundaries. In repeated simulations
with this and similar A-tract fragments, the static curvature emerged
spontaneously and it became more evident as the chain length
increased \cite{Mzjacs:00}. To obtain a reference non-A-tract DNA, we
have re-shuffled manually base pairs of the A-tract repeat. We
preferred this randomized sequence to commonly used GC-rich straight
fragments in order to keep the base pair content identical and reduce
the noise that could cause small variations in gel mobility and hide
the subtle differences we were going to detect.

\subsection*{Spontaneous Development of Curvature in Simulations}

All four trajectories exhibited stable dynamics with DNA structures
close to the B form. Table \ref{Tenpa} shows parameters of the final
1ns-average conformations. They all have remarkably similar
helicoidals corresponding to a typical B-DNA. For example, the average
helical twist estimated from the best-fit B-DNA experimental values
\cite{Kabsch:82} gives $34.0\pm 0.2$\degree\ and $33.8\pm 0.2$\degree\
for the A-tract fragment and the randomized sequence, respectively.
At the same time, the rms deviations from the canonical structures
vary more significantly.

\begin{figure}
\centerline{\psfig{figure=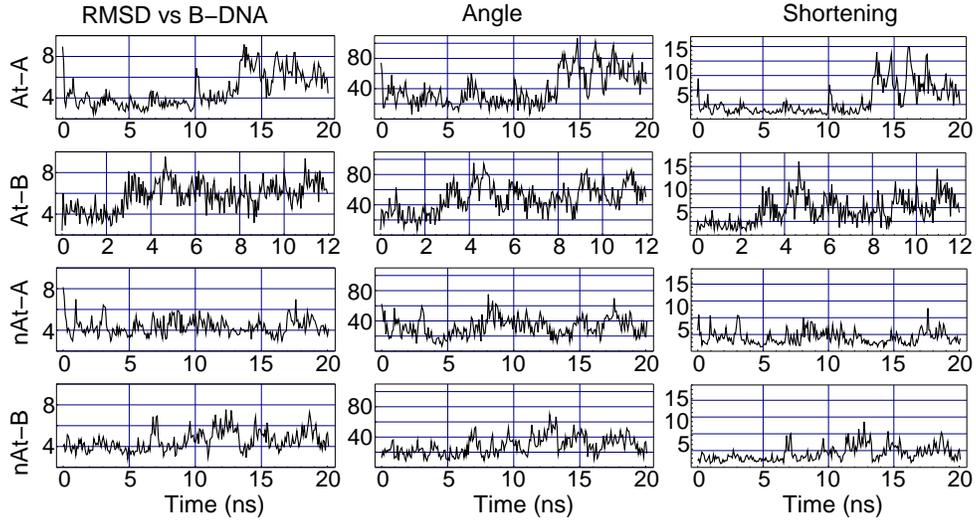,width=15cm,angle=0.}}
\caption{\label{Ftcbnd} The time variation of some parameters that
characterize the overall DNA shape. The plates are grouped in raws for
the same trajectory and in columns for the same parameter. The first
column displays the non-hydrogen atom rmsd (in angstr{\"o}ms) from the
fiber canonical B-DNA \pcite{Arnott:72}. The second column shows the
bend angle in degrees. The last column shows the shortening, that is
the excess length of the curved DNA axis with respect to its
end-to-end distance. For example, 10\% shortening means that the
end-to-end distance is 10\% shorter than the curved trace. The traces
were smoothed by averaging with a window of 75 ps in At-B and 150 ps
otherwise. }\end{figure}

As shown in Fig. \ref{Ftcbnd}, during the first few nanoseconds, the
rmsd from the canonical B-DNA quickly leveled at around 4 \AA\ in all
four trajectories. For the A-DNA start this corresponds to a rapid
transition to B-form with reduction of rmsd from the initial 10.7 \AA.
The subsequent dynamics is remarkably different for the At and nAt
trajectories. In At-A and At-B, after some delay, the rmsd
value drastically increased and stabilized at a higher level of around
6 \AA. The traces of the bend angle and the axis shortening indicate
that this was a transition to a significantly larger curvature. In
contrast, for nAt trajectories, Fig.  \ref{Ftcbnd} exhibits only
fluctuations at roughly the same level as in At-A and At-B before the
transition.

\onecolumn
\begin{table*}[t]\caption{\label{Tenpa} Some structural parameters
of standard and computed DNA conformations. The helicoidals are the
sequence averaged values computed with program Curves \pcite{Curves:}.
All distances are in angstr{\"o}ms and angles in degrees. }
\begin{tabular}[t]{|ddddddd|}
           & Xdisp
             & Inclin
               & Rise
                 & Twist
                   & RMSD vs A-DNA
                     & RMSD vs B-DNA\\
\hline
A-DNA & -5.4 & +19.1 & 2.6 & 32.7 &  0.0 & 10.7 \\
B-DNA & -0.7 & -6.0  & 3.4 & 36.0 & 10.7 & 0.0  \\

At-A  & +0.1 & -4.0 & 3.5 & 34.2 & 11.6 & 5.9   \\
At-B  & -0.4 & -5.2 & 3.5 & 34.5 & 11.6 & 6.8   \\
nAt-A & -0.1 & -4.2 & 3.5 & 34.3 & 10.6 & 3.8   \\
nAt-B & -0.1 & -4.7 & 3.5 & 34.4 & 11.2 & 4.1   \\
\end{tabular}
\end{table*}

\begin{figure}
\centerline{\psfig{figure=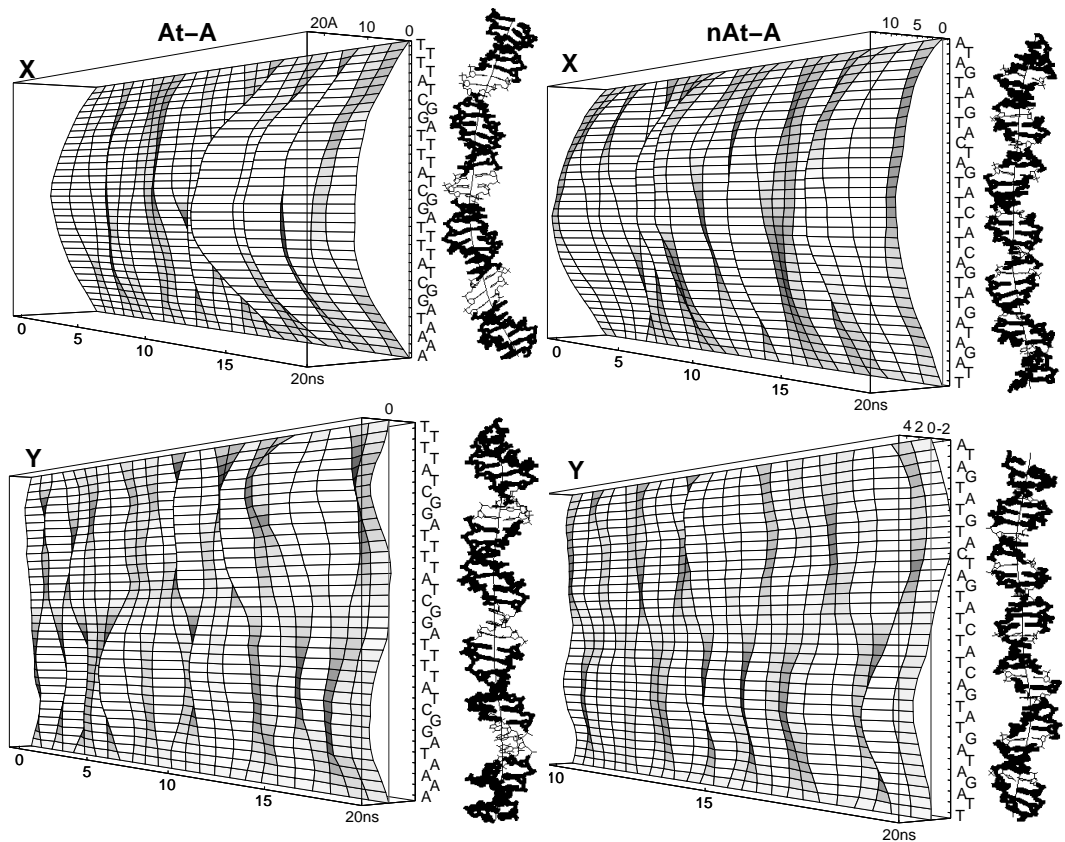,width=15cm,angle=0.}}
\caption{\label{Fbndsr}
The time evolution of the overall shape of the helical axis in At-A
and nAt-A. The axis of the curved double helix is computed as the best
fit common axis of coaxial cylindrical surfaces passing through sugar
atoms, which gives solutions close to those produced by the Curves
algorithm \pcite{Curves:}. The two surface plots labeled $\bf X$ and
$\bf Y$ are constructed by using projections of the curved axis upon
the XOZ and YOZ planes, respectively, of the global Cartesian frame
shown in Fig. \pref{Ftcdc}a. Any time section of these surfaces gives
the corresponding projection averaged over a time window of 400 ps.
The horizontal deviation is given in angstr\"oms and, for clarity, its
relative scale is two times increased with respect to the true DNA
length. Shown on the right are the corresponding views of the final
1ns-average conformations. The AT base pairs are shown by thicker
lines.}\end{figure}
\twocolumn

The origin of this difference is analyzed in Fig.
\ref{Fbndsr}. It displays dynamics of the overall DNA shape by using
two orthogonal projections of the helical axis. A planar bend would
give a plane in the $\bf Y$ projection and a curved surface in the
$\bf X$ projection. A sharp increase of curvature in At-A after the
13th nanosecond is evident.  Analogous event occurred in At-B after
about 3 ns. In agreement with Fig. \ref{Ftcbnd}, the two nAt surfaces
show fluctuations with amplitudes similar to those during the first 13
ns of At-A. This pattern probably corresponds to a generic type of
dynamics characteristic of arbitrary 35-mer DNA fragments.

\begin{figure}
\centerline{\psfig{figure=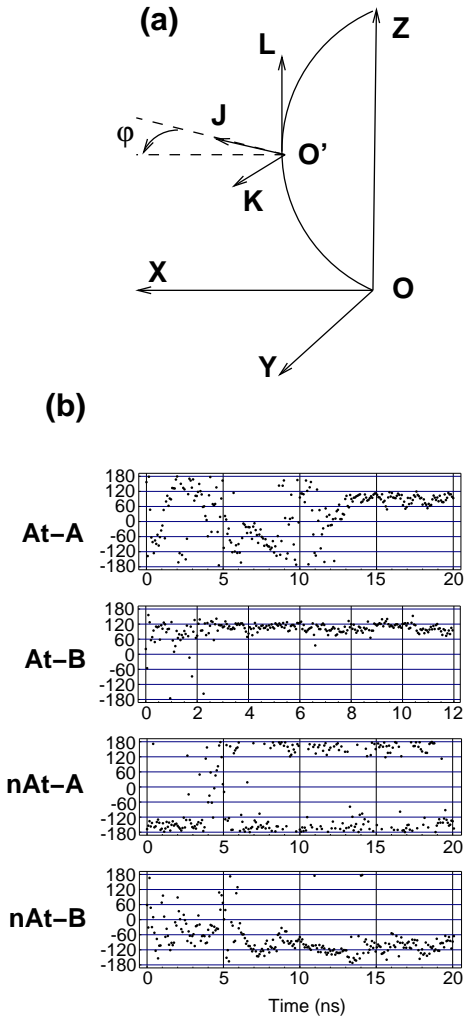,width=7.5cm,angle=0.}}
\caption{\label{Ftcdc}
(a) Geometric constructions used for evaluating the DNA bending.  The
two coordinate frames shown are the global Cartesian coordinates
(OXYZ), and the local frame constructed in the middle point of the
curved DNA axis according to the Cambridge convention (O'JKL)
\pcite{Dickerson:89}. The curve is rotated with two its ends fixed at
the Z-axis to put the middle point in plane XOZ. The bending direction
is measured by angle $\varphi$ between this plane and vector $\bf J$
of the local frame. By definition, this vector points to the major DNA
groove along the short axis of the reference base pair
\pcite{Dickerson:89}. Consequently, the zero $\varphi$ value
corresponds to the overall bend towards the minor groove in the middle
of the DNA fragment. (b) The time evolution of the bending direction
as measured by the $\varphi$ angle in plate (a) (in degrees). The
traces have been smoothed by averaging with a window of 75 ps in At-B
and 150 ps otherwise.}\end{figure}

Comparison of the three columns of plots in Fig. \ref{Ftcbnd}
indicates that fluctuations usually occurred simultaneously in all
three parameters, which means that the bending dynamics makes a
major contribution to the rmsd from B-DNA. Its values shown in Table
\ref{Tenpa} are actually much larger than would be for straight
conformations with the same helical parameters.  For instance, the
rmsd between the At-A and At-B structures in Table \ref{Tenpa} was 2.3
\AA\ only because, as we show below, they were bent in the same
direction.

\subsection*{Convergence of Trajectories}

A DNA molecule with detectable static curvature can either have a
minimum of potential energy in a bent state or its energy valley should
have a special shape such that a bent form has larger conformational
entropy \cite{Olson:93}. In both cases such state represents a
free energy minimum where MD trajectories should be trapped. The
question is, however, how long a real MD trajectory should stay in a
bent conformation to be representative. Some experiments suggest that
bending dynamics in DNA fragments of only 100 base pairs may involve
relaxation times longer than a microsecond \cite{Song:90,Okonogi:99},
therefore, no practical procedure exists to prove rigorously that
computed conformations are representative. Nevertheless, if several
trajectories converge to the same state from very different starting
points, one can argue that this state is an attractor in the
conformational space, which is a necessary condition of the static
curvature. The reciprocal convergence of trajectories starting from
canonical A- and B-DNA, therefore, is a very important aspect of these
simulations. The minimal B-DNA model is not expected to give stable
A-DNA structures and we did not try to equilibrate the initial A-DNA
states. The start form the A-form is important because it provides an
independent dynamic assay with a very different entry to the B-DNA
family, which allows one to verify convergence of trajectories to
specific conformations. We analyze separately two levels of structural
convergence.

\subsubsection*{Overall shapes}

The rmsd comparison between At-A and At-B is shown in Fig.
\ref{F2Drmsd}. It clearly demonstrates that At-A and At-B trajectories
managed to come very close to each other even though their starting
points were significantly separated in conformational space. The
initial rmsd of 10.7 \AA\ between the canonical 35-mer A- and B-DNA
forms eventually went down to as low as 1.3 \AA. The final fall of the
rmsd occurred when the curvature drastically increased (compare Figs.
\ref{Ftcbnd} and \ref{F2Drmsd}). Moreover, during the last nanoseconds
the bending direction was virtually identical in At-A and At-B and
essentially fixed at around 90\degree\ (see Fig. \ref{Ftcdc}b), which
explains the origin of the black rectangle in the upper right corner
of Fig. \ref{F2Drmsd}. This direction corresponds to bending towards
the minor groove at approximately three base pair steps from the
middle GC pair (see Fig. \ref{Ftcdc}a), that is at the 3' end of the
third A-tract in Fig.  \ref{Falseq}.

The nAt trajectories exhibited qualitatively different features. The
rmsd comparison of any two long intervals of nAt-A and nAt-B gives
fluctuations between 3 and 6 \AA\ without any clear time trend.
Figure \ref{Ftcbnd} shows that the rmsd from B-DNA also fluctuated
between 3 and 6 \AA\ and that it correlated with bending parameters.
As seen in Fig. \ref{Fbndsr} the molecule really was not straight.
According to Fig. \ref{Ftcdc}b the bending directions in nAt-A and
nAt-B were well defined but slightly different. They neither diverged
nor converged, remaining at around 100\degree\ from each other. The
molecule shows no signs of slow straightening, which would give a
decrease of fluctuations in Fig. \ref{Ftcbnd} and an increase in
scattering of directions in Fig. \ref{Ftcdc}b. All this suggests that
bent shapes are favored over straight ones, and that there are many
stable bends, with transitions between them being too rare to be
sampled by our simulations.

\subsubsection*{Groove profiles and local structures}

Dynamics of the minor groove profiles is shown in Fig. \ref{Fmgkt}.
There are evident qualitative resemblance as well as some subtle
differences between these four surfaces. In At-B, the characteristic
regular groove shape has established early, with significant widenings
in the three zones between the A-tracts. The very left widening is
somewhat different probably because it occurs between anti-parallel
A-tracts. In At-A, the profile strongly changed at the beginning, but
also established by the end of the 10th nanosecond. Although the final
At-A and At-B profiles are not identical, they are clearly similar,
with good correspondence of local widenings and narrowings.

The two nAt surfaces show little similarity between each other, but
qualitatively their shapes are not very different from those for the
A-tract fragment, with modulations of similar wavelengths and
amplitudes. This looks somewhat counter-intuitive because, in
experiments, regular oscillations of the minor groove widths are
observed only in A-tract repeats \cite{Burkhoff:87}, and this
structural periodicity is certainly related to that of the sequence.
However, such behavior is exactly what one should expect if the waving
of the backbone results from its intrinsic compression. In this case,
the groove modulations should occur regardless of the base pair
sequence and their characteristic wave lengths should be determined by
the backbone stiffness as well as overall helical pitch and diameter.
This explains why the waves in the left-hand and the right-hand plates
in Fig. \ref{Fmgkt} have roughly similar scales, even though only the
A-tract sequence is periodical. In experiment, however, such
modulations can be observed only if their phases are fixed in time,
which is the case of A-tract repeats. For random sequences, like the
one we use as a reference, the fine structure should be smoothed out
on averaging over the whole ensemble.

Figure \ref{FB1B2} compares $\rm B_I/B_{II}$ backbone dynamics in the
two At trajectories. There are many similarities in dynamics as well
as in the final configurations. The convergence is better near both
ends and within A-tracts. The dissimilar distributions of the
conformers in the middle corresponds to the difference in minor groove
profiles in Fig. \ref{Fmgkt}. In A-tracts, the $\rm B_{II}$ conformers
are very rare in T-strands and tend to alternate with $\rm B_I$ in
A-strands. Figure \ref{Fhlpa} compares local helical parameters in the
last average structures. Only the Buckle and Propeller traces exhibit
large scale modulations phased with the helical screw. All parameters
strongly fluctuate and these fluctuations are apparently chaotic with
rather dissimilar phases in the two structures.

\subsubsection*{Coupling between the levels}

\begin{figure}
\centerline{\psfig{figure=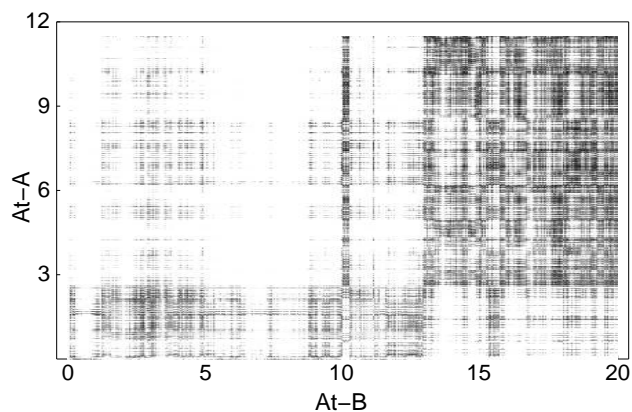,width=10cm,angle=0.}}
\caption{\label{F2Drmsd}
A 2D density plot of the rms difference between At-A and At-B.
Conformations spaced by 2.5 ps intervals were first averaged over of
50 and 25 ps intervals in At-A and At-B, respectively, and the resulting
structures compared between the trajectories. Darker shading implies
smaller rmsd values. The lower left corner corresponds to the initial
structures, that is the canonical A- and B-forms, with rmsd about 10.7
\AA. The shaded rectangle in the upper right corner demonstrates
convergence of the two trajectories to the same bent state. The values
above 4 \AA\ are not shaded whereas in the darkest zones it falls down
to 1.3 \AA. The black vertical band at approximately 10 ns indicates
that At-A shortly visited the final state 3 ns before the definite
transition.}\end{figure}

\begin{figure}
\centerline{\psfig{figure=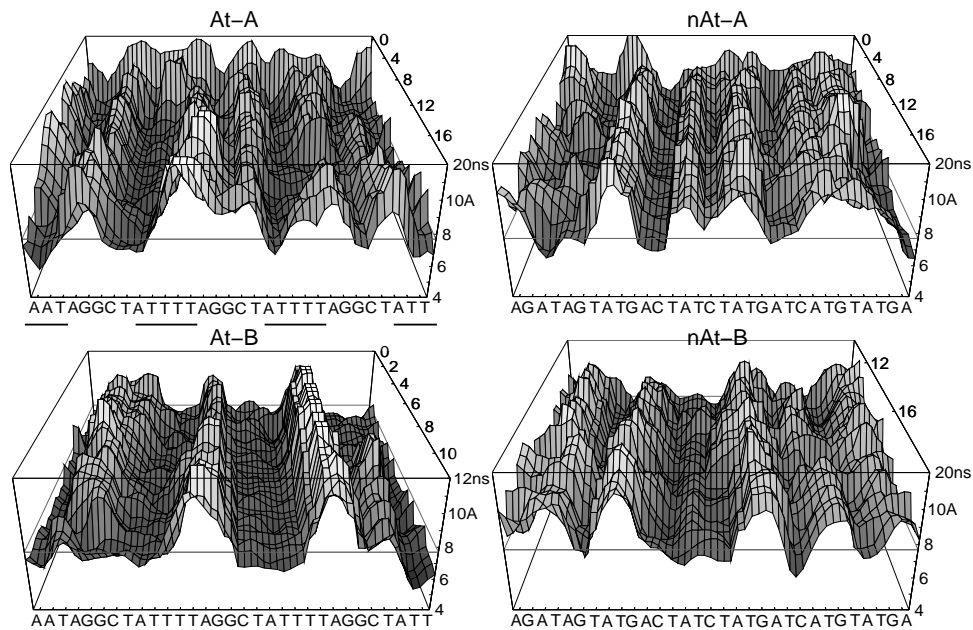,width=14cm,angle=0.}}
\caption{\label{Fmgkt}
The time evolution of the profile of the minor groove in the four
trajectories. The surface plots are formed by time-averaged successive
minor groove profiles, with that on the front face corresponding to
the final DNA conformation. The groove width is evaluated by using
space traces of C5' atoms \pcite{Mzjmb:99}. Its value is given in
angstr\"oms and the corresponding canonical B-DNA level of 7.7 \AA\ is
marked by the thin straight lines on the faces of the box. The
sequences are shown for the corresponding top strands in Fig.
\pref{Falseq} with the 5'-ends on the left. The A-tracts are
underlined. Note that the groove width can be measured only starting
from the third base pair from both termini. }\end{figure}

\begin{figure}
\centerline{\psfig{figure=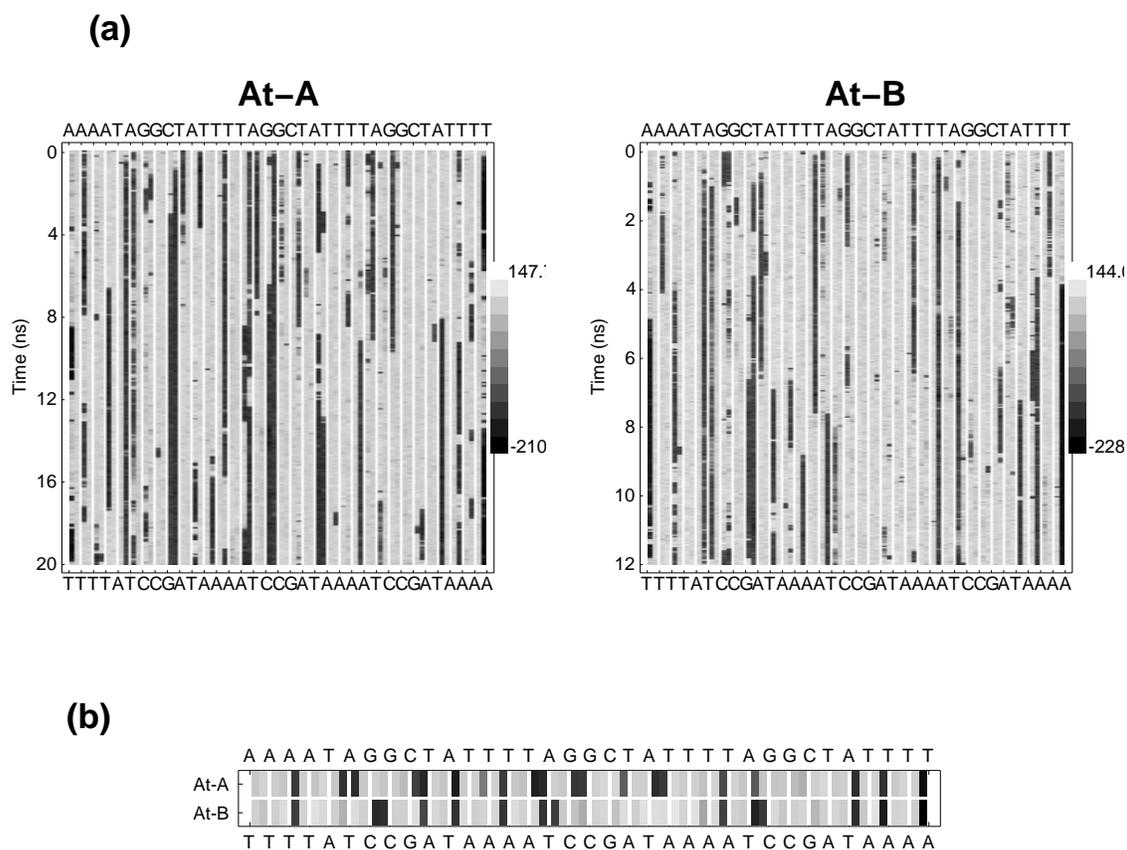,width=16cm,angle=0.}}
\caption{\label{FB1B2}
(a) Dynamics of $\rm B_I$ and $\rm B_{II}$ backbone conformers in At-A
and At-B.  The B$_{\rm I}$ and B$_{\rm II}$ conformations are
distinguished by the values of two consecutive backbone torsions,
$\varepsilon$ and $\zeta$. In a transition, they change concertedly
from (t,g$^-$) to (g$^-$,t). The difference $\zeta -\varepsilon$ is,
therefore, positive in B$_{\rm I}$ state and negative in B$_{\rm II}$,
and it is used as a monitoring indicator, with the corresponding gray
scale levels shown on the right. Each base pair step is characterized
by a column consisting of two sub-columns, with the left sub-columns
referring to the sequence written at the top in 5'-3' direction from
left to right.  The right sub-columns refer to the complementary
sequence shown at the bottom. (b) Comparison of the final
distributions of $\rm B_I\ and\ B_{II}$ backbone conformers in At-A
and At-B shown in the same way as in plate (a). }\end{figure}

\begin{figure}
\centerline{\psfig{figure=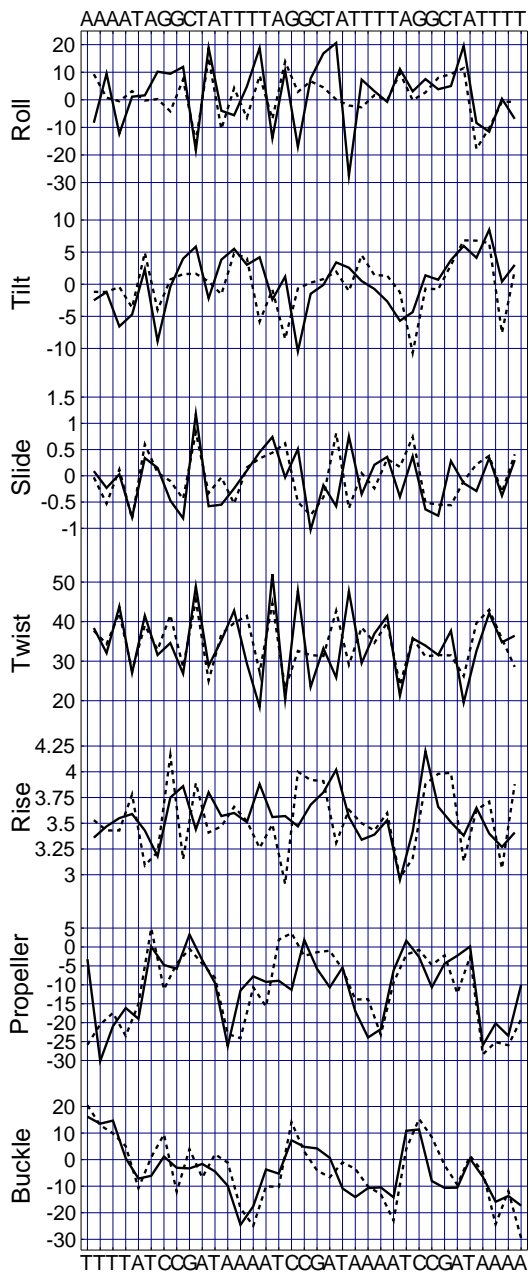,height=18cm,angle=0.}}
\caption{\label{Fhlpa}
Sequence variations of helicoidal parameters in the last 1ns-average
structures of At-A and At-B.  The sequence of the first strand is
shown on the top in 5' -- 3' direction. The complementary sequence of
the second strand is written on the bottom in the opposite direction.
All parameters were evaluated with the Curves program \pcite{Curves:}
and are given in degrees and angstr\"oms. At-A -- solid line, At-B --
dashed line.  }\end{figure}

Figures \ref{F2Drmsd} and \ref{Ftcdc} demonstrate that At-A and At-B
arrived at the same statically bent state. This dynamics contrasts
those of the two nAt trajectories and it strongly suggest that the
curved DNA shape of the A-tract fragment is an attractor of
trajectories with a meta-basin of attraction comprising both canonical
A and B DNA forms. Figures \ref{Fmgkt}-\ref{Fhlpa} show that the
bending convergence is accompanied by some clear trends in local
conformational dynamics. These local features are probably coupled to
bending, however, a close look reveals that this coupling is very
loose. The convergence of the minor groove profiles in Fig.
\ref{Fmgkt} is at best qualitative. Figure \ref{FB1B2} indicates that
active backbone dynamics continued after the curvature has established
and that one can pick up rather different distributions of conformers
from the ensemble of bent structures. The noisy traces in Fig.
\ref{Fhlpa} obtained by averaging over two similarly bent ensembles
suggest that the helical parameters are far from being constant. The
natural conclusion follows that convergence of the bending dynamics
does not require unique specific local conformations, i. e.  that the
bent state is microheterogeneous.

\onecolumn

\twocolumn

\subsection*{The Magnitude and The Character of Bending in the A-tract
Repeat}

The experimental magnitude of bending caused by A-tracts was earlier
estimated by several groups with different approaches
\cite{Levene:86,Calladine:88,Ulanovsky:86,Koo:90}. The reported bend
angles were between 11\degree\ and 28\degree\ per A-tract, and
18\degree\ is presently considered as the most reasonable estimate
\cite{Crothers:99}. The curvature somewhat varies with the base pair
sequence and depends upon environmental conditions such as the
temperature, the concentration of counterions etc. Although in
calculations all these details cannot yet be properly taken into
account a quantitative comparison with experiment is instructive.

When the curvature has established, that is after 13 ns of dynamics in
At-A and after 3 ns in At-B, the bend angle oscillated around
60\degree (see Fig. \ref{Ftcbnd}). In the consecutive 1ns-averaged
conformations its value was between 42\degree\ and 74\degree, with the
average of 54\degree\ for 16 such structures. This value corresponds
to 54/4=13.5\degree\ per A-tract, that is close to the lower
experimental estimate. A larger value of 54/3=18\degree\ results, however,
if one assumes, as suggested by some experimental observations
\cite{Dickerson:94,Young:95} that the A-tracts are straight,
and that the bending actually occurs in the three zones between them.
Yet another estimate is obtained from the increase of bending with
respect to the shorter 25-mer fragment studied earlier
\cite{Mzjacs:00}. It appears that one additional A-tract and junction
zone increase the overall bend by 20-22\degree. We see that the
magnitude of bending in simulations is rather close to experimental
estimates, and that the agreement is better if the curvature is really
localized in the junction zones between A-tracts.

\begin{figure}
\centerline{\psfig{figure=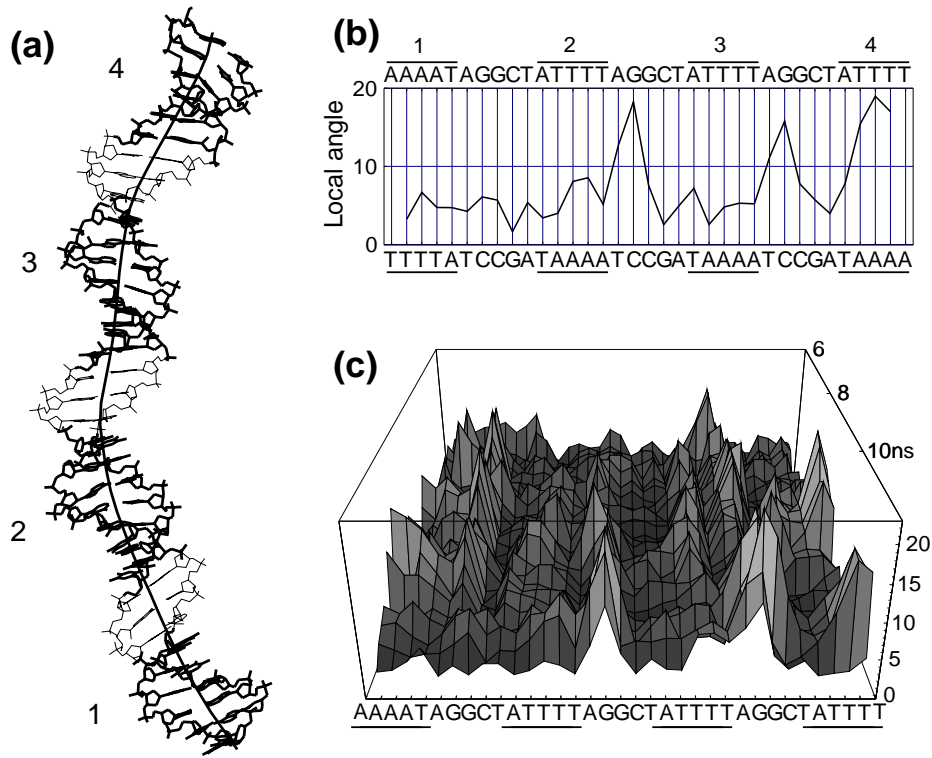,width=14cm,angle=0.}}
\caption{\label{Flcbnd}
(a) The last 1-ns average structure of At-B shown in the XOZ
projection according Fig. \pref{Ftcdc}a. The AT base pairs are
highlighted.
(b) The quantified distribution of curvature in the structure shown in
plate (a). The local bending angle is evaluated by moving a sliding
window along the helical axis. The window size was 3 base pair steps,
with the measured values assigned to its center. The sequences of
strands are given as in Fig. \pref{Falseq} with the A-tracts
underlined and numbered. (c) Dynamics of local bending in At-B. The
surface plot is formed by time-averaged successive profiles like that in
plate (b), with the front face of the box corresponding to the end
of the trajectory.}\end{figure}

Figure \ref{Flcbnd} presents a closer look at how the local curvature
is distributed in the last 1-ns average structure of At-B. The total
bending angle is about 50\degree. Three zones contribute more than
other to the overall bend. The two junctions between A-tracts 2, 3 and
4 are bent in an identical direction which is close to that of the
whole structure. Together they contribute around 40\degree\ to the
total bend, which is the largest local positive contribution. In
contrast, the strongly curved fourth A-tract makes a negative
contribution because its direction diverges by more than 90\degree.
The third A-tract is virtually straight. Finally, A-tracts 1 and 2 and
the junction zone between them exhibit a smooth curvature with a
stable ``good'' direction and contribute the remaining 20\degree\ of
the total bend.

\onecolumn
\twocolumn

The foregoing analysis certainly is not free from pitfalls. For
instance, the apparent smooth curvature can result from time averaging
of several alternative local bends. Nevertheless, Fig. \ref{Flcbnd}
indicates that there are zones in this DNA fragment that are bent more
than other and that two such zones are distinguishable between
A-tracts. Figure \ref{Flcbnd}c displays the local bending dynamics in
At-B. It is seen that the main features noticed in plates (a) and (b)
were quite visible during the whole trajectory. Moreover, the zone
between the first two A-tracts also sometimes carried an increased
curvature. However, it would be incorrect to conclude that A-tracts
are straight. They just exhibit generally smaller and distributed
curvature than the junction zones. This curvature is usually directed
towards the minor groove, therefore, it does not cancel out in
averaged structures.

The foregoing pattern agrees qualitatively with the recent NMR
\cite{MacDonald:01} and X-ray data \cite{Chiu:99} as well as the
character of bending earlier observed in calculations
\cite{Young:98,Sprous:99}. Many earlier reported X-ray structures of
A-tracts suggested that they produce an intrinsically straight DNA
compared to other sequences \cite{Young:95}. Our calculations do not
contradict these observations because the crystal A-tract structures
should be additionally straightened due to special crystallization
conditions \cite{Sprous:95,Dlakic:96b,Dickerson:96}, and because a
single short A-tract may in fact be somewhat less curved than that
inserted in a long DNA fragment.

\begin{figure}
\centerline{\psfig{figure=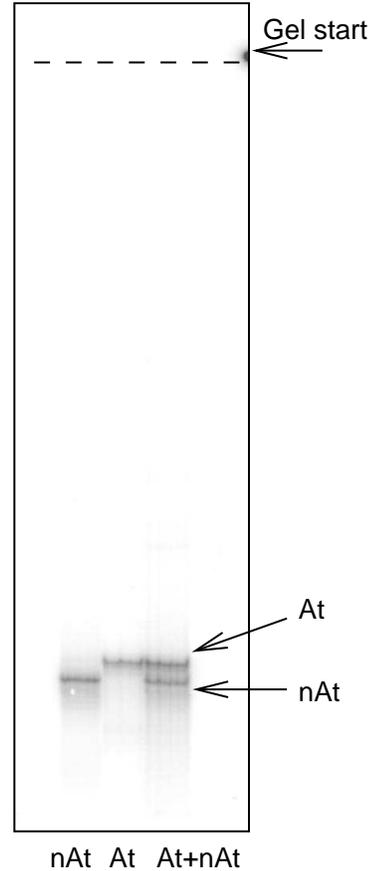,width=6cm,angle=0.}}
\caption{\label{Fgel}
Gel mobility assay. The two $^{32}$P-labeled 35 bp DNA constructs
(At and nAt) were electrophoresed in 16\% plyacrylamide gel buffered with
Tris-borate, pH 8.6. The gel was dried and autoradiographed. The
lanes labeled At, nAt, and At+nAt correspond to the A-tract
repeat, the random sequence, and their mixture, respectively. Bands
assigned to each DNA fragment are marked by arrows. }\end{figure}

\subsection*{Verification of Curvature by Gel Electrophoresis}

The sequence induced static DNA curvature was first noticed owing to
reduced migration rate of curved DNA fragments in gel electrophoresis
\cite{Marini:82}. Later gel migration studies provided a wealth of
information on curvature in A-tract repeats \cite{Crothers:92}. The
difference in gel mobility between straight and curved DNA rapidly
grows with chain length, therefore, the curvature was usually studied
in rather long DNA fragments. Data for sequences shorter than 50 bp
are rare \cite{Diekmann:87c}, and, to our best knowledge, it has never
been shown that curved and straight 35-mers could be distinguished.
Nevertheless, subtle sequence effects in double stranded oligomers of
around 10 bp were detected with higher gel concentration
\cite{Chen:88}, and one could hope that this would work for somewhat
longer sequences as well. Alternatively, the effect of bending could
be enhanced by inserting constructed fragments into a long stretch of
straight DNA \cite{Wu:84}, but this would complicate further analysis
because the DNA molecules used in experiments and calculations could
no longer be identical.

Figure \ref{Fgel} shows comparison of the acrylamide gel mobility
of these two fragments. As expected, the A-tract repeat exhibits a
reduced rate of migration. The difference is quite significant so that
the two molecules are well resolved both in separate lanes and when
mixed in the same sample. Owing to the identical base pair content,
the minor factors such as the number of tightly bound counterions and
water molecules is reduced here to the possible minimum and, most
probably, the observed difference is entirely due to the curvature in
the A-tract fragment.

\section*{Discussion}

\subsection*{Comparison with Earlier Studies}

To our knowledge, the only earlier successful unbiased simulations
aimed at reproducing A-tract induced curvature in DNA have been
reported by David Beveridge group \cite{Young:98,Sprous:99}. These
simulations were carried out in full water environment with explicit
counterions. The character of the phased A-tract bending appeared
oscillatory with a period of at least 3 to 4 ns \cite{Young:98}.
Because duration of trajectories was only 5 ns, it was difficult to
confirm the static character of bending and distinguish between
essential and occasional observations. Therefore, conclusions
concerning applicability of different models were not restrictive and
left room for many theories. Our simulations have the same goal and a
similar set up, but we use a simpler model system. The primarily long
term interest in B-DNA models with implicit or semi-implicit
representation of environment is connected with approximate
simulations of very long DNA molecules \cite{Cheatham:00}.  As shown
here the minimal model can also capture, at least qualitatively,
important sequence effects like the A-tract induced curvature.

Several features in our calculations correspond well to those observed
earlier, notably, spontaneous development of quasi-sinusoidal minor
groove profiles in both A-tract and non-A-tract sequences and strong
bends in junction zones between A-tracts. In contrast to earlier
simulations, however, the curvature here emerged after several
nanoseconds of dynamics and the difference between the A-tract and
non-A-tract structures did not reduce with time. One should note also
that the A-tract structures computed with the minimal model are very
close to experimental data as regards the helical pitch and the
absolute groove sizes \cite{Mzjacs:00,Mzjcc:01}. In standard AMBER and
CHARMM simulations, B-DNA always appears somewhat underwound and the
narrowest A-tract minor grooves remain 1-2 \AA\ wider than in
experimental structures \cite{Young:98,Sprous:99,Strahs:00}. The
origin of this subtle bias remains unclear, and attempts to reduce it
have been made in the very recent modifications of the AMBER force
field \cite{Cheatham:99,Wang:00}. In the minimal model, this bias was
compensated by fitting reduced phosphate charges, which apparently
improved the A-tract structures and stabilized the curvature. 

A significant enforcement of the present results compared to our
previous reports \cite{Mzjacs:00,Mzjbsd:01} consists in the direct
comparison of bending {\em in silico} and {\em in vitro}, which became
possible owing to increased length of the DNA fragment. Such a
possibility is rather unique for MD simulations and we believe this
approach presents considerable interest for future studies. Sequence
effects in DNA fragments of 35-50 base pairs can be probed in both MD
simulations and gel electrophoresis. Such experiments are rapid and
inexpensive, which progressively becomes the case for MD simulations
as well.

\subsection*{Comparison with Theories of DNA Bending}

The origin of intrinsic curvature in DNA remains unclear. Theories
that explain it always assume some specific balance of interactions in
the DNA structure, and that is why these theories are perhaps more
important than the particular role of A-tracts. The list of available
interactions is well-known, but the question is which of them is the
driving force. Below we briefly analyze our results in contexts of
some theories.

\subsubsection*{Base Pair Stacking Models}

According to any mechanism that starts from base pair stacking, like
the wedge or the junction models \cite{Trifonov:80,Levene:83,Wu:84},
a curved DNA must be built out of asymmetric blocks, with their
structures determined by base pair sequence. The bending, therefore,
must be accompanied by repetition of local structures in identical
sequence fragments. This fundamental theoretical prediction fails for
the static bends observed here, which confirms earlier conclusions
\cite{Mzjacs:00,Mzjbsd:01}. The structures of sequence repeats in the
bent state are microscopically heterogeneous and convergence to
specific local conformations is not necessary for bending. As shown
above, the A-tract trajectories arrive at a single bent state, but the
minor groove profiles in Fig.  \ref{Fmgkt} are only similar, not
identical as well as local helical parameters and backbone
conformations in Figs. \ref{FB1B2} and \ref{Fhlpa}.

\subsubsection*{Counterion Electrostatic Models}

An alternative model that recently attracted much attention considers
solvent cations trapped in A-tract minor grooves as the initial cause
of bending \cite{Hud:99}. The role of counterions in this phenomenon
is rather controversial \cite{McFail-Isom:99,Chiu:99,McConnell:00},
and a few general comments are necessary before considering our
results. Because straight DNA structures correspond to symmetric
minima of electrostatic energy bends can result from symmetry breaking
in the charge distribution, namely, if positive external charges
accumulate at one DNA side it should bend towards them
\cite{Mirzabekov:79,Strauss:94,Young:97a,Rouzina:98,Hud:99}. However,
the same situation is well interpreted by other models of bending.
Namely, in a curved double helix, the phosphate groups at the inner
edge must approach, which creates regions of low potential that should
be populated by counterions if they are available \cite{Levene:86}. In
the first case the counterion-DNA interactions are sequence specific
and they cause bending. In the second case they are structure specific
and they stabilize pre-existing curvature.

Two physically different models of sequence-specific counterion
involvement can be distinguished. In the first one the counterions act
locally. When a counterion is placed in one of the DNA grooves between
two phosphate groups their electrostatic interaction becomes
attractive, which narrows the groove \cite{Rouzina:98}. As in some earlier
models \cite{Drew:85,Burkhoff:87}, the global curvature
results from a general mechanical link between groove deformations and
bending. In contrast, the second model is purely electrostatic. Here
the minor grooves of A-tracts act as flexible ionophores
\cite{Hud:99,Hud:01} and trap counterions. Since in phased sequences
they occur at the same DNA side the double helix bends towards them to
relax the long range phosphate repulsion at the opposite side.

The second model employs the general idea initially proposed for
protein DNA interactions \cite{Mirzabekov:79} and confirmed
experimentally for free DNA \cite{Strauss:94}. However, it
qualitatively disagrees with a cornerstone experimental observation
concerning the A-tract induced bending, namely, that an A-tract can be
characterized by a definite bend angle regardless of its length and the
distance from other A-tracts. When the length of an A-tract
exceeds one helical turn both sides of the double helix appear
neutralized. As a result, the curvature should decrease in the series
$\rm (A_{12}N_9)_n\ -\ (A_{14}N_7)_n\ -\ (A_{16}N_5)_n$ because the
length of the non-neutralized N-tracts is reduced, and furthermore, in
sequence $\rm (A_{16}N_5)_n$ the bend angle per A-tract should be
drastically reduced with respect to that in $\rm (A_6N_5)_n$, for
example, because the distance between the repulsive N-tracts is
increased. These predictions apparently disagree with the experimental
trends \cite{Haran:89} although additional experiments are perhaps
necessary to check them.

The first model cannot explain the origin of the A-tract curvature
because only multivalent counterions can cause significant bends
\cite{Rouzina:98} whereas bending is commonly observed in buffers
containing EDTA and other chelating agents. Also, the optimal counterion
position for this type of bend is at the entrance of the groove and
not inside, therefore, it cannot be both strong and sequence specific.
The last argument agrees with the recent MD studies of correlations
between the minor groove width and positioning of counterions.
Notably, there is no such correlation when only counterions
interacting with bases are considered \cite{McConnell:00}.  In
contrast, a correlation exists for counterion positions at the groove
entrance \cite{Hamelberg:01}. The last observation corresponds to the
structure specific binding better than to the sequence specific one.
Structure-specific interactions can explain all experimental results
concerning the preferential binding of counterions in A-tracts
\cite{Hud:97,Hud:99,Stellwagen:01}, which makes such data
intrinsically neutral as regards different models of bending.

In our calculations all counterion effects are considered
non-specific, and the results obtained indicate that modulations of
DNA grooves and static bending are physically possible without
breaking the charge symmetry around DNA. Although simulations alone
cannot prove the real mechanism all experimental and computational
observations taken together suggest that solvent counterions are
hardly responsible for the intrinsic curvature in DNA, which by no
means questions their important role in DNA structure and function.

\subsubsection*{Compressed Backbone Theory}

The main idea of the compressed backbone theory \cite{Mzjacs:00} was
outlined in Introduction. All seemingly paradoxical MD observations
from which it originally emerged are confirmed here. Notably, this
theory predicts that, with any base pair sequence, the backbone
stiffness should cause smooth modulations of DNA grooves. The helical
symmetry becomes broken with the base pair stacking perturbed, which
creates regions of intrinsic curvature. In a ``random'' DNA, the local
curvature changes its direction with time because groove widenings and
narrowings migrate slowly along the double helix.  As a result, the
generic DNA appears straight on average although it is curved locally.
In sequences where certain base pair properties strongly alternate,
the phases of backbone oscillations appear fixed.  In this case the
local curvature can sum up to give macroscopic static bends, as in
A-tract repeats. This theory considers a macroscopically curved DNA
as an ``idioform'' characterized by topological attributes, rather
than a structure with fixed atom positions. These attributes are the
bend direction and the phase of groove modulations.  The
microheterogeneity of the bent state should be expected because the
same waving backbone profile is compatible with many alternative local
conformations.

A competition between the stacking interactions and the backbone
compression postulated by this theory is characteristic of physical
systems called frustrated \cite{Liebmann:86}.  Consider the common
textbook example of three anti-ferromagnetic spins in a triangle
configuration. The optimal orientation of each pair is anti-parallel,
but all three pairs cannot be anti-parallel in a triangle. There is
always at least one parallel pair and the ground state appears
degenerate. Now consider a circular duplex DNA with a homopolymer
sequence. The compressed backbone causes groove modulations, but there
are no preferable regions for narrowings and widenings and the ground
state appears strongly degenerate. The similarity between these two
examples is evident. One usual physical consequence of frustration is
very important for biology, namely, the possibility of a glassy state
where microscopic transitions are dramatically slowed down.
Transitions between wavy backbone configurations in a long DNA can be
very slow because many groove narrowings and widenings must be moved
concertedly. This may explain observations of supra-microsecond
relaxation times in bending dynamics of relatively short DNA fragments
\cite{Song:90,Okonogi:99}.

Finally, the compressed backbone theory offers a new view of some
environmental effects upon the curvature. Common physical factors like
the temperature, counterions, and various dehydrating agents are
long-known to change slightly the helical pitch of DNA
\cite{Depew:75,Anderson:78,Lee:81}, which can reasonably be attributed
to the dependence of the state of the DNA backbone upon the solvent
screening of phosphates. These same factors modulate significantly the
sequence specificity of nucleases probably by changing the shape of
DNA grooves \cite{Drew:84}, and produce complex effects upon the
intrinsic curvature
\cite{Diekmann:85,Diekmann:87c,Marini:84,Sprous:95,Dlakic:96b,Jerkovic:00}.
It seems wise to postpone any detailed interpretation of these facts
for future studies, but one can just note that with intrinsic
frustration outlined above a very small change in the partial specific
backbone length can induce significant global changes in the DNA
structure.

\subsection*{Possibilities of Experimental Verification of Backbone
Compression}

The the compressed backbone theory does not give simple rules for {\em
a priori} calculation of curvature in any sequence. Nevertheless, it
offers some predictions that can be checked in experiments. It
suggests, for example, that the A-tract curvature can be relaxed by
introducing single-strand breaks. It seems interesting also to examine
the possible relationship between the backbone compression and supercoiling.
There is a consensus that intrinsic bends affect the shape of the
superhelical DNA \cite{Laundon:88,Yang:95}. Unlike other models, however,
the compressed backbone theory predicts that the intrinsic curvature
should vary under superhelical stress in a rather special way.
Namely, with a positive density, the backbone is stretched and the
curvature of an internal A-tract repeat should be reduced. Conversely,
the curvature should increase when the superhelical density is
negative. Diekmann and Wang earlier observed that the A-tract
structure changes under superhelical stress \cite{Diekmann:85}, and
their approach may serve for a more specific experimental verification
of the above predictions.

\end{document}